\documentclass{acm_proc_article-sp}

\usepackage{xspace}
\usepackage{graphicx}
\usepackage{grffile}
\usepackage{enumitem}
\usepackage{pifont}
\usepackage{algorithmic}
\usepackage{algorithm}
\usepackage{float}
\usepackage{moreverb}
\usepackage{booktabs}
\usepackage{multirow}
\usepackage{subfigure}
\usepackage{color}
\usepackage{etoolbox}
\usepackage{booktabs}
\usepackage{arydshln}
\usepackage{url}
\usepackage{times}

\graphicspath{{.}{./figure/}}

\newcommand{\term}[1]{\emph{#1}}
\newcommand{\xref}[1]{Section~\ref{#1}}

\newcommand{\fref}[1]{Figure~\ref{#1}}

\newcommand{\ie}{i.\,e., \@}
\newcommand{\eg}{e.\,g., \@}

\newcommand{\etal}{et~al.\xspace}
\newcommand{\perc}{\,\%\xspace}

\begin{document}

\title{Multi-source Multipath HTTP (mHTTP): A Proposal}

\author{
\alignauthor \large{Juhoon Kim, Ramin Khalili, Anja Feldmann}
\and
\makebox[\linewidth]{\email{\rm \{jkim,ramin,anja\}@net.t-labs.tu-berlin.de}, \affaddr{\rm Telekom Innovation Laboratories / TU Berlin, Germany.}}
\vspace*{0.3cm}
\and
\alignauthor \large{Yung-Chih Chen, Don Towsley}
\and
\makebox[\linewidth]{\email{\rm \{yungchih,towsley\}@cs.umass.edu}, \affaddr{\rm University of Massachussets, Amherst, USA.}}
}

\maketitle

\thispagestyle{empty}

\section{Abstract}
\label{sec:abstract}

Today, most devices have multiple network interfaces. Coupled with wide-spread
replication of popular content at multiple locations, this provides substantial
path diversity in the Internet. We propose Multi-source Multipath HTTP, mHTTP, 
which takes advantage of all existing types of path diversity in the Internet. 
mHTTP needs only client-side but not server-side or network modifications as it 
is a receiver-oriented mechanism. Moreover, the modifications are restricted to 
the socket interface. Thus, no changes are needed to the applications or to the 
kernel.

As mHTTP relies on HTTP range requests, it is specific to HTTP which accounts for 
more than 60\% of the Internet traffic~\cite{MAIER09-ODC}. We implement mHTTP
and study its performance by conducting measurements over a testbed and in the
wild. Our results show that mHTTP indeed takes advantage of all types of path 
diversity in the Internet, and that it is a viable alternative to Multipath 
TCP\footnote{Multipath TCP is an extension to regular TCP that allows a user 
to simultaneously use multiple interfaces for a data transfer~\cite{RFC-MPTCP}.} 
for HTTP traffic. mHTTP decreases download times for large objects 
up to $50\%$, whereas it does no harm to small object downloads.

\section{Introduction}
\label{sec:intro}

\begin{figure*}[ht]
    \begin{center}
	\begin{minipage}[h]{0.4\linewidth}
        \begin{center}
		\subfigure[Regular HTTP/TCP architecture at a client.]{\includegraphics[width=0.8\linewidth,natwidth=610,natheight=642]{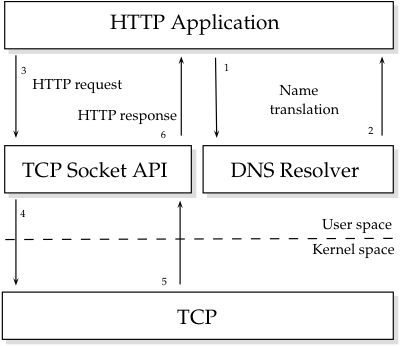}}
        \end{center}
	\end{minipage}
	\hspace{1pt}
	\begin{minipage}[h]{0.4\linewidth}
        \begin{center}
		\subfigure[mHTTP architecture at a client.]{\includegraphics[width=0.8\linewidth,natwidth=610,natheight=642]{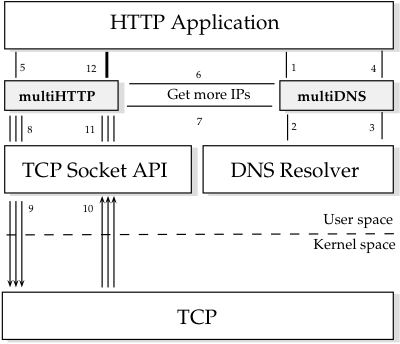}}
        \end{center}
	\end{minipage}
	\caption{\label{fig:arch} Structural differences between HTTP/TCP and mHTTP}
   \end{center}
   \vspace*{-0.3cm}
\end{figure*}

In today's Internet, one of the main detriments in user experience is
completion times of data transfers that for large objects is limited by network
capacity. However, recent developments have opened new opportunities for
reducing end-to-end latencies. First, most end-user devices have multiple
network interfaces (e.g., 3G/LTE and WiFi interfaces for smart-phones). Second,
popular contents are often available at multiple locations in the network. When
combined, these provide substantial path diversity within the Internet that
can be used by users to improve their quality of experience.

Previous work has taken partial advantage of this path diversity in the
Internet. Multipath TCP (MPTCP) uses the path diversity available between a
single server and a single client~\cite{RFC-MPTCP,RAICIU12-HHC}. Application
specific download managers are other examples of related work that benefits
from the path diversity between a single server and a single
client~\cite{URL-FLASH,URL-JDOWN}. Content Distribution Networks (CDNs) provide
replication of content and smart matching of users to appropriate CDN server,
e.g., via PaDIS~\cite{POESE10-ICD} or ALTO~\cite{RFC-ALTO} services, which
takes advantage of this replication. Moreover, there are application specific
video streaming protocols that try to take advantage of the replication of streaming contents
provided by CDNs~\cite{KASPA12-PHD,ADHIKARI-UNU,TIAN12-TAA}. Bittorrent is
another sophisticated application which takes advantage of content replication
among its users~\cite{COHEN03-IBR}. The drawback of each of the above
approaches is that they do not utilize all of different types of path diversity in
the Internet or if they do, they are application specific.

We propose Multi-source Multipath HTTP, mHTTP, which enables users to establish
simultaneous connections with multiple servers to fetch a single content.
mHTTP is designed to combine the advantage obtained from distributed network
infrastructures provided by CDNs with the advantage of multiple interfaces at 
end-users. Unlike existing proposals: a) mHTTP is a purely receiver-oriented 
mechanism that requires no modification either at the server or at the network,  
b) the modifications are restricted to the socket interface; hence, no changes 
are needed to the applications or to the kernel, and c) it takes advantage of 
all existing types of path diversity in the Internet.

mHTTP is proposed for HTTP traffic, which accounts for more than $60\%$ of the
total traffic in today's Internet~\cite{MAIER09-ODC}. As stated in
Popa~\etal~\cite{defacto-http}, HTTP has become the de-facto protocol for
deploying new services and applications. This is due to the explosive growth of
video traffic and HTTP infrastructure in the Internet in recent years.
mHTTP is primarily designed to improve download times of large file transfers 
(such as streaming contents). Measurements results have shown that connections 
with large file transfers are responsible for the bulk of the total volume of 
traffic in the Internet~\cite{MAIER08-ENS}. Furthermore, while mHTTP decreases 
download times for large objects by up to $50\%$, it does no harm to small 
object downloads as shown in Section~\ref{sec:evaluation}.

The key insight behind mHTTP is that HTTP allows chunking a file via byte range
requests and that these chunks can be downloaded from different servers as long
as these servers offer the identical copies of the object\footnote{The extra
header added by the application might be different from server to server. mHTTP
parser, refer to Section \ref{sec:multihttp}, deals with the application
headers.}. mHTTP learns about the different servers that host the same content
by either using multiple IP addresses returned by a regular DNS query, sending
multiple queries to multiple DNS servers, or utilizing the eDNS
feature~\cite{RFC-EDNS}. It also works with a single server when
multiple paths are available between the receiver and the server. Hence, mHTTP
can also be used as an alternative to MPTCP for HTTP traffic. 

The mHTTP design consists of: (i) multiHTTP: a set of modified socket APIs
which splits a content into multiple chunks, requests each chunk via individual
HTTP range requests from the available servers, reassembles the chunks and
delivers the content to the application. (ii) multiDNS: a modified DNS resolver 
that obtains IP addresses for a server name by harvesting the DNS replies and/or by
performing multiple lookups of the same server name by contacting different
domain name servers.

Our key contribution is the concept of mHTTP along with a prototype implementation 
and evaluation. We evaluate the performance of mHTTP through measurements over a 
testbed and in the wild. We compare the performance of mHTTP with regular HTTP 
operating over single-path TCP and MPTCP. We observe that
\vspace*{-0.3cm} 
\begin{itemize}
\item mHTTP indeed takes advantage of all types of path diversity in the Internet. 
\item For large object downloads, it decreases download times up to $50\%$
compared to the single-path HTTP transmission. Moreover, it does no harm to small 
object downloads.
\item mHTTP performs similar to MPTCP while only requiring receiver-side 
modifications. As MPTCP requires changes to the kernel, both at the sender 
and receiver, we consider mHTTP to be a viable alternative when running HTTP. 
\end{itemize}
\vspace*{-0.3cm}
The comparison with MPTCP is performed in a single-server scenario as MPTCP is 
restricted to the use of a single server. mHTTP, on the other hand, can be used 
both in single-server and in multi-server scenarios.  

This paper is structured as follows. In the next section, we provide an
overview of mHTTP from a system viewpoint. We give a detailed description of
our prototype implementation in Sections~\ref{sec:multidns},
\ref{sec:multihttp}, and \ref{sec:schedule}. In Section \ref{sec:evaluation},
we study the performance of mHTTP through measurements. Related work is
presented in Section \ref{sec:related}. Section \ref{sec:discussion} provides a
summary of our results and discusses next steps.

\section{M\lowercase{ulti-source} M\lowercase{ultipath} HTTP}
\label{sec:mhttp}

Content Distribution Network (CDN) provide wide-spread replication of popular 
content at multiple locations in the Internet. Multi-source Multipath HTTP (mHTTP) 
is designed to combine the advantage obtained from such a diversity with the advantage of
diverse network connectivity at end-users. In this section, we describe the high-level concept of mHTTP.

\subsection{Regular HTTP over TCP}
Before we discuss the design of mHTTP, we review various components of HTTP
communication over TCP. As illustrated in~\fref{fig:arch}(a), when an HTTP
application tries to download an object from a web server, it first requests to
the local ~\term{DNS resolver} to translate the human-friendly URL to a set of
IP addresses. And then it sends an HTTP request to establish a connection to one 
of these addresses. ~\term{TCP Socket API} is the interface to the underlying 
transport protocol (TCP). The~\term{TCP stack} in the kernel ensures reliable data
transmission and congestion control. Note that (i) one domain name may be
associated with multiple IP addresses. Moreover, (ii) different DNS servers may
return different IP addresses. This is often observed in server infrastructures
which serve popular contents. (i) occurs when the load is spread across
multiple servers~\cite{POESE10-ICD}. (ii) occurs in Content Delivery Networks
(CDNs). However, even though the content is in principle available at multiple
locations, traditional HTTP/TCP cannot take advantage of this. To overcome this 
limitation, we propose a novel protocol, mHTTP, built on the regular HTTP-over-TCP 
architecture.

\subsection{mHTTP}
mHTTP is designed with the following three key features in mind:
\vspace*{-0.3cm}
\begin{itemize}
    \item mHTTP must take advantage of multiple built-in interfaces,
    multiple paths, and multiple data sources, by establishing simultaneous
    connections via multiple interfaces to multiple data servers where the
    identical content is stored.
    \item mHTTP must not make any modifications on the server-side
    infrastructure or the protocol stack.
    \item The client-side implementation must be transparent to the application,
    i.e., modifications must be limited to only socket APIs.
\end{itemize}
\vspace*{-0.3cm}

\begin{figure}[t]
   \begin{center}
        \includegraphics[width=0.8\linewidth]{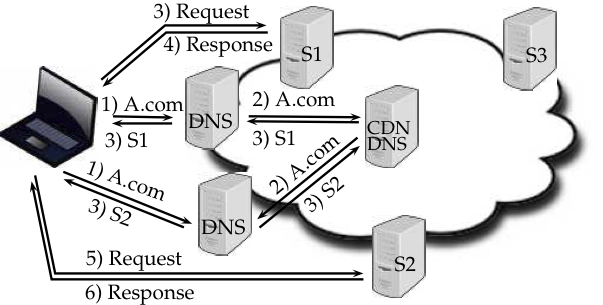}
    	\caption{An mHTTP, with two connections over two available interfaces, operation in a CDN. Servers S1, S2, and S3 host replicas of content of A.com.}
        \label{fig:mhttp-cdn} 
    \end{center}
    \vspace*{-0.3cm}
\end{figure}

The key idea of mHTTP is to use the HTTP range request feature to fetch
different content chunks from different servers. We define a chunk as a block
of content delivered within one HTTP response message. mHTTP includes two
components, \term{multiHTTP} and \term{multiDNS} as shown in~\fref{fig:arch}(b). 
These components extend the functionality of HTTP and the DNS resolver.
The main purpose of multiHTTP is to handle chunked data delivery between the
application and multiple servers; and that of multiDNS is to collect IP
addresses of available content sources.~\fref{fig:mhttp-cdn} illustrates the
process for a $2$-connection mHTTP session in a CDN.  

multiHTTP is the core component of mHTTP. It is responsible for the management
of data chunks; taking advantage of multiple content servers and therefore of path
and network diversity; and scheduling chunk requests. multiHTTP intercepts all
messages sent from the application to the remote end-host (e.g., server). When
a TCP connection is identified as an HTTP connection, on reception of an HTTP
request from the application, the multiHTTP module modifies the header of the HTTP
request by adding a range field. The HTTP response header includes the file
size. Thus mHTTP can issue multiple range request to multiple servers
that serve the same object and their IP addresses are known via multiDNS. 

If the connection is not an HTTP connection, mHTTP falls back to regular socket
APIs. Also, for an HTTP connection, if the server does not reply with a partial
data response to the HTTP request, or if multiDNS only returns a single IP
address and the client is single homed mHTTP may still decide to fall back to
regular HTTP.

multiDNS obtains different IP addresses by performing multiple lookups of the
same server name by contacting different DNS servers (\ie{the local DNS servers
of the upstream ISPs for each of its interfaces, a Google DNS server, an OpenDNS
server, etc.}). Additionally, it can use the eDNS extension to uncover many more
servers in a CDN infrastructure.

\section{Implementation of \lowercase{multi}DNS}
\label{sec:multidns}

\begin{figure}[ht]
  \begin{center}
	\begin{minipage}[h]{0.8\linewidth}
		\subfigure[Top-1000 domain names]{\includegraphics[width=\linewidth]{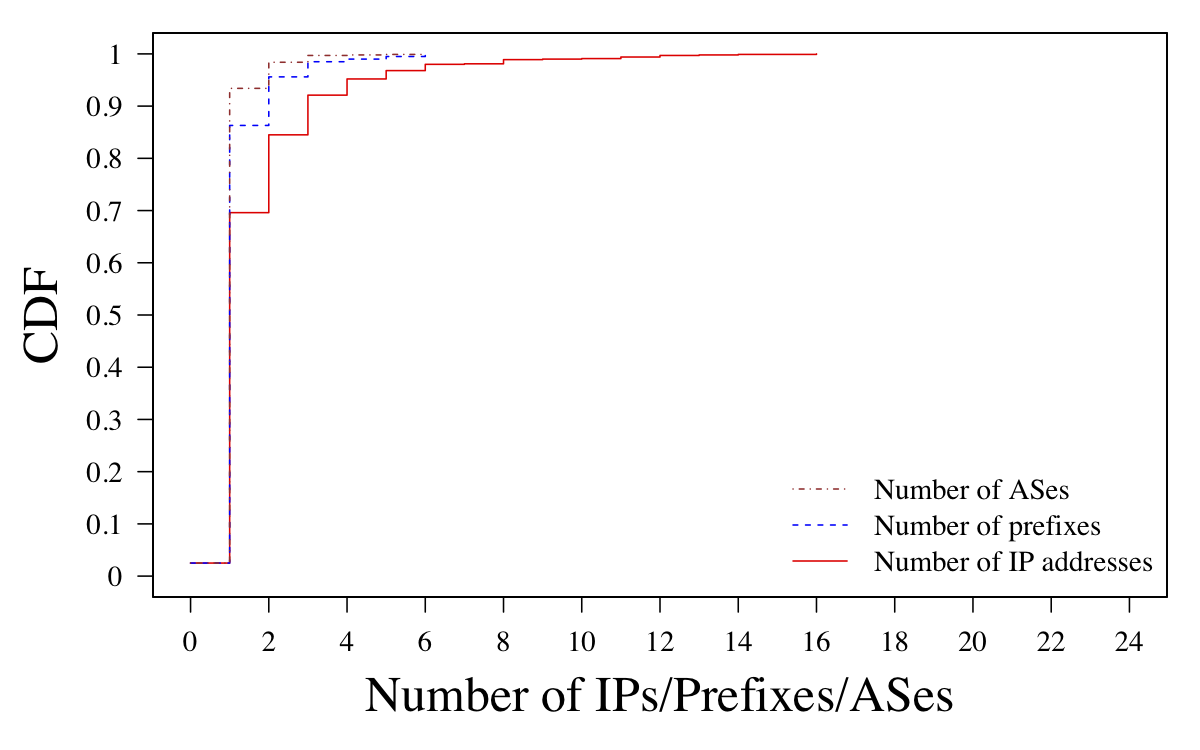}}
	\end{minipage}
	\end{center}
	\hspace{1pt}
	\begin{center}
	\begin{minipage}[h]{0.8\linewidth}
		\subfigure[Top-300 domain names]{\includegraphics[width=\linewidth]{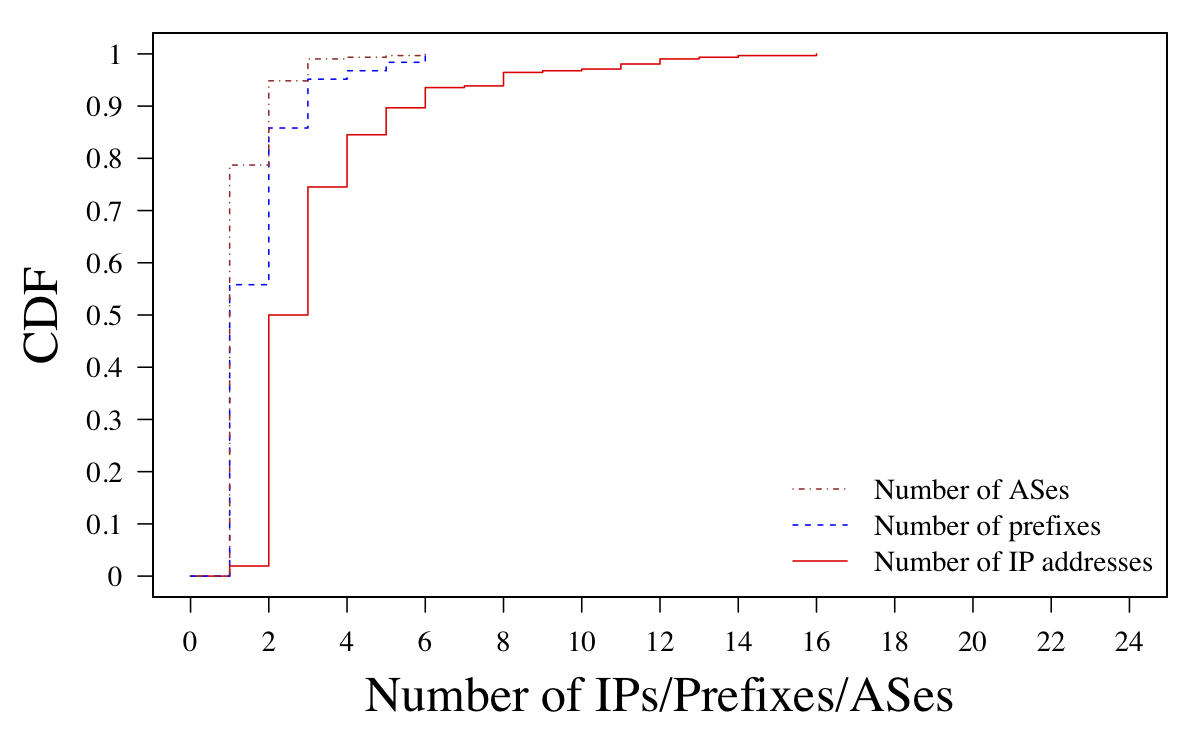}}
	\end{minipage}
	\end{center}
	\caption{\label{fig:alexa} CDF of IP addresses, prefixes, and AS numbers for top-1000/top-300 domain names obtained from a single query to the local DNS.}
  \vspace*{-0.3cm}
\end{figure}

In this section, we describe multiDNS the core element designed for discovering  
IP addresses of servers that hold replicas of a content. Note that mHTTP also 
works with a single server when multiple paths are available between the
receiver and the server (refer to Section \ref{sec:mhhtp_vs_mptcp} for an example). 

\subsection{Data Source Diversity}
Before we discuss details of the multiDNS implementation, we analyze how many
IP addresses we receive from a single query for a hostname. We choose the top-1000 hostnames
provided by~\term{Alexa.com} and request the resolution of these hostnames by sending 
DNS queries to the local DNS server of a client residing in a university campus. As illustrated
in~\fref{fig:alexa}(a), even with a single query to the local DNS server, approximately 
30\perc of the total hostnames are associated to more than one IP address and respectively 
10\perc and 5\perc of the total hostnames are in different network prefixes and reside 
in different ASes (Autonomous Systems). When performing two lookups, one query to the local DNS and another to the Google DNS 
(not shown as a figure), these numbers increase to 35\perc (IP addresses), 17\perc 
(prefixes), and 7\perc (ASes). This can be seen as an evidence that CDNs may provide 
a different set of IP addresses to a user depending on the choice of a DNS server. 
More evidence of the content diversity can be found in the work of 
Poese~\etal~\cite{POESE11-IGD}.

In~\fref{fig:alexa} (b), we narrow down the scope to the top 300 hostnames and our
result shows that almost all hostnames are associated with at least two IP addresses.
Given the fact that the major fraction of the total traffic originates from a
small number of popular content providers~\cite{AGER10-RCT,BRESLAU99-WCZ} and
the fact that the top 15 domains account for 43\perc of the total HTTP traffic
in a large European ISP~\cite{MAIER09-ODC}, the fraction of the traffic
contributed by providers through multiple servers should be significant.  

\subsection{Getting an IP address by a Hostname}
When an application needs to obtain an IP address from a human-readable
URL, it invokes name resolvers such as~\term{gethostbyname()} or
\term{getaddrinfo()}. The resolver, then, creates a request message and sends it
to the local DNS server usually provided by the local ISP. Depending on the content
sources, if content is only available at a single server, the DNS returns the
IP address of that particular server so that the request can be routed to the
server. However, if content is available at multiple places (e.g., a server 
farm or CDNs), DNS returns a list of IP addresses. In the case of multiple IP addresses, 
a typical behavior of an application is to choose the first IP address in order 
to establish the connection and to discard the rest. multiDNS, however, keeps the 
rest of the IP addresses for later use.

\subsection{Getting more IP addresses}
As mentioned above, different DNS servers may provide different sets of IP
addresses. Therefore, it is worthwhile querying multiple
DNS servers in order to obtain more IP addresses. multiDNS plays the
role of managing the identities of different resolver of different access networks,
whenever an interface is activated and the IP address is assigned. It also
handles the DNS query by validating the availability of a local DNS server in each
access network for each interface\footnote{The local DNS information is
collected when a DHCP request is completed}. If local DNS is still available 
at the point of a name translation, a query to that content is made to the local DNS of that
particular access network. For each interface, multiDNS receives a list of IP
addresses from each access network, and chooses desired number of IP addresses
from every list. Hence, if the desired contents are available at CDNs, mHTTP does not 
only retrieve them from the CDNs accessible to the public, but from the 
CDN nodes in the CDN server farm known to the local DNS resolvers.

\section{Implementation of \lowercase{multi}HTTP}
\label{sec:multihttp}

The main task of multiHTTP is to interpret mHTTP for regular HTTP speakers such
as web servers and client-side applications. 


\subsection{HTTP Byte Range Request}
RFC2616~\cite{RFC-2616} specifies the use of a \term{byte range request} which
enables the partial delivery of content. A client initiates such a request by
adding a \term{range} field within the header of an HTTP request message
including offsets of the first byte and the last byte of the partial content.
If the server supports this operation, it replies with 206 as the status code
(on acceptance of the request message) followed by sequences of bytes.
Otherwise, the server replies with a different status code (e.g., 200 OK on
success). Note that a block of partial content is referred to as a chunk in
this paper. Although \cite{RFC-2616} defines this operation as an optional
feature, our tests on well-known web servers during the development of mHTTP
show that almost all web servers accept \term{range} requests.

Partial content delivery is widely employed by HTTP-based downloaders in order
to continuously resume fetching a transferred file. Another common
usage of this feature is multi-threaded downloading implemented in some
software, i.e., \cite{URL-FLASH} and~\cite{URL-JDOWN}. Such software boosts
download speed by fetching different parts of the content over different
connections using the partial content delivery. At first sight, mHTTP is
similar to those software; however mHTTP operates in the Socket API thus 
helps existing HTTP software to utilize the bandwidth more effectively. 
As mHTTP is designed to communicate with multiple servers containing 
identical copies of the content, it is clearly distinguished from 
multi-thread and downloader approaches.

\subsection{mHTTP Buffer}

\begin{figure}[t]
    \begin{center}
	\includegraphics[width=0.8\linewidth]{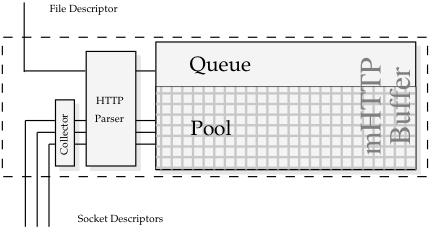}
	\caption{\label{fig:multihttp} multiHTTP design: a) \term{mHTTP buffer} stores out-of-order received data b) \term{HTTP parser} examines and modifies HTTP headers c) \term{collector} gathers data from individual TCP connection buffers.}
    \end{center}
    \vspace*{-0.3cm}
\end{figure}

multiHTTP initializes \term{mHTTP buffer} and creates a file descriptor
associated with the buffer when \term{socket()} is called by the application.
The buffer consists of a \term{queue} and a \term{pool} of content blocks
(chunks). The queue is a large memory block that is continuously read by the
application. Thus, the file descriptor plays the role of a communication
channel between the application and the mHTTP buffer (\fref{fig:multihttp}).
The pool maintains multiple content blocks in which chunked data collected from
individual TCP buffers is stored. A content block can be indexed by the
combination of the socket descriptor and the starting byte of the chunk. Data
within content blocks is moved to the queue as soon as it is continuous from
the last byte that is stored in the queue. The size of the queue does not
grow greatly since it is continually drained by the application. However, the size of
the pool needs to be sufficient to store out-of-order received chunks. We study the 
required size of the mHTTP buffer through measurements in~\xref{sec:evaluation}.
If mHTTP decides to fall back to the regular HTTP, or if the connection is not
an HTTP connection, one of the socket descriptors replaces the file descriptor
and the mHTTP buffer is discarded.

\subsection{Manipulating HTTP Headers}
Once the connection is identified as an HTTP connection, multiHTTP enables an
HTTP parser, which examines HTTP messages during the content delivery period.
The tasks of the HTTP parser are mainly three-fold:

\begin{itemize}
\item \textbf{HTTP request manipulation} The HTTP parser adds the \term{range}
field to the end of the header with the specified chunk size when the initial
HTTP request is sent by the application. When the response message to the
initial request arrives back to the application, multiHTTP knows the size of
the file and whether or not the server accepts a byte range request.
\item \textbf{Parsing HTTP headers} The HTTP parser extracts and stores
important information from the request and response headers such as
availability of content, support for the partial content delivery, content
size, and the byte range of the content block.
\item \textbf{Response header management} In order to allow applications to use
mHTTP without modification, the behavior of mHTTP must be the same as that of a regular
HTTP communication from an application's point of view. To this end, the HTTP
parser replaces the initial response header (206) with a header that indicates
the acceptance of the request (200). All subsequent headers are discarded by
the HTTP parser.
\end{itemize}

\subsection{Connections}
Upon confirmation of the complete delivery of the initial response message,
multiHTTP establishes additional TCP connections using different IP addresses
provided by multiDNS. In order to obtain another IP address, multiHTTP invokes
\term{get\_ip()} from multiDNS (see 6 and 7 in~\fref{fig:arch}). The mechanism 
used by multiDNS to selects IP addresses is independent of the operation of 
multiHTTP. The current version of multiDNS hands IP addresses over to multiHTTP 
in the order that they are retrieved. Similarly, the number of connections to be 
used is configurable.

multiHTTP operates \term{collector}, a background process that collects data
from individual TCP connection buffers. Each new connection must be
attached to the \term{collector} as soon as it is successfully established.
Likewise, a connection can be detached from the \term{collector}.

Determining what content chunk to request over each connection is another
important task of multiHTTP. It keeps track of the requested chunks and
decides which chunk to ask on the next request message after the previous chunk
on the same connection is completely fetched. The mechanism used for such
decision is further explained in the next section.


\section{Scheduling}
\label{sec:schedule}
\begin{figure}[t]
    \begin{center}
    \includegraphics[width=0.8\linewidth]{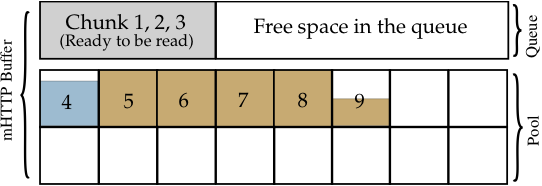}
    \caption{A bottleneck in mHTTP buffer. Chunk $4$ is downloaded over a slow
    connection. Chunks $5$, $6$, $7$ ,$8$ are downloaded over a fast
    connection. These chunks cannot proceed to the queue before chunk $4$ is
    completely received. A scheduler is needed to better allocate chunks across
    different connections.}
    \label{fig:bottleneck}
    \end{center}
\end{figure}

Different connections may have different qualities, in terms of latency, capacity,
and loss rate. This may cause reordering of the chunks received at the mHTTP
buffer. \fref{fig:bottleneck} illustrates an example of such reordering. We 
have two connections: one slow and one fast (in terms of download time). The $4$th chunk
is downloaded over the slow connection and the 5th, 6th, 7th, and 8th chunks are
downloaded over the fast connection. As the download of the $4$th chunk is not
finished yet, these later chunks cannot be moved to the queue. Hence, a mechanism
that allocates chunks to different connections plays a critical role in
multiHTTP. In this section, we explore mHTTP's design choice with regard to
chunk \term{scheduling}.

For simplicity, we assume a client with two interfaces (e.g.
$e_0$ and $e_1$). Let $S_0 = \{s_{01}, s_{02},\cdots, s_{0N1}\}$ and $S_1 =
\{s_{11}, s_{12},\cdots, s_{1N1}\}$ be respectively sets of servers available
to the client through $e_0$ and $e_1$. Note that $S_0$ and $S_1$ are not
necessarily disjoint sets. $N_1$ and $N_2$ are numbers of discovered servers
over $e_0$ and $e_1$. Let $P_0$ and $P_1$ denote the connections established through
$e_0$ and $e_1$ to servers in $S_0$ and $S_1$. In this paper, we limit the
number of established connections over each interface to one and the total number of
connections to two. However, our implementation can accept an arbitrary number of 
connections per interface. 

We measure the instantaneous throughput of each connection by measuring the number
of bytes received at the mHTTP every $20$ ms. We use a moving average to
estimate the average throughput of each connection: 
$$\overline{THR}_{new} = 0.8 * THR + 0.2 * \overline{THR}_{old}$$ 
where $THR$ is the instantaneous throughput measured every $20$ms and
$\overline{THR}$ is the estimated average throughput.
We denote by $\overline{THR}_0$ and $\overline{THR}_1$ the estimated average 
throughput measured over connections $P_0$ and $P_1$.

Let $L$ denote the size of the object and $C$ the chunk size, both measured 
in bytes. We denote by $N = \left\lfloor L/C \right\rfloor + 1$ the number 
of chunks to be fetched by the client and by $1, 2, 3, \cdots, N$ the chunk 
numbers. Here, $\left\lfloor x \right\rfloor$ is the largest integer not 
greater than $x$. Also, let $D$ be the set of chunks that have not yet 
been requested for download. $D$ is a sorted set based on the chunk numbers.  

The scheduling algorithm decides what chunk to request over each connection. For
example, if a chunk is successfully fetched over connection $P_0$, the next chunk
over this connection must be carefully chosen in order to avoid a bottleneck
situation such as presented in~\fref{fig:bottleneck}. mHTTP decides the next 
chunk over $P_0$ uses the following mechanism:
\begin{enumerate}
	\item calculate $T_0 = \left\lfloor \max{(\overline{THR}_0,\overline{THR}_1)} / \overline{THR}_0 \right\rfloor$;
  \item ask for the $T_0$'th chunk from the set $D$. If $T_0 > |D|$, no
    chunk is requested. $|D|$ is the size of set $D$. 
	\item Remove the requested chunk from the set $D$.  
\end{enumerate}
$T_0$ predicts the number of chunks that can be delivered over the best connection 
among $P_0$ and $P_1$ while one chunk is transmitted over $P_0$. If $P_0$ is the best 
connection, then $T_0=1$. When mHTTP needs to issue a new request over $P_0$, it does 
not request the next chunk but skips to the $T_0$'th chunk from the set $D$. 

mHTTP uses similar mechanism to decide the next chunk to be requested over connection 
$P_1$, with the modification that $T_0$ is replaced with 
$T_0 = \left\lfloor \max{(\overline{THR}_0,\overline{THR}_1)} / \overline{THR}_1 \right\rfloor$

In~\xref{sec:evaluation}, we compare the performance of our scheduler with 
a baseline that multiHTTP simply requests the next chunk in $D$, whenever it 
needs to issue a new chunk request over a connection. We show that our scheduler 
can efficiently reduce the the size of the mHTTP buffer without affecting the 
performance of mHTTP.

\section{Performance Evaluation}
\label{sec:evaluation}

In this section, we study the potential benefit of using mHTTP in different
indoor and outdoor scenarios through measurements. We study how mHTTP takes
advantage of different types of diversity in the Internet and compare its
performance to that of regular HTTP operating over single-path TCP and MPTCP. 

We use the download completion time as the performance metric in our
evaluation. It is defined as the duration between the first SYN packet from the
client and the last data packet from the servers. The download completion times are
measured for different file sizes,~\ie $4$MB, $16$MB, and $64$MB. We run each
measurement $30$ times and show the median, $25-75\%$ percentiles (boxes), and
dispersion (lines, $5-95\%$ percentiles). In each round of measurement, we
randomize the configuration sequence to account for traffic dependencies and/or
correlation from time to time and from size to size. Specifically, we randomize
the order of file sizes, the choice of protocol (e.g., single-path, mHTTP, and
MPTCP), and the choice of chunk sizes for mHTTP. 

The servers run an \term{Apache2} web server on port 80 and hold copies of the same 
files. The client uses \term{wget} in order to retrieve the files from the servers
and runs on the Linux operating system with the kernel version 3.5.7. We
use $10$ MSS as the initial size of the congestion window. Furthermore, TCP 
\term{Cubic}~\cite{cubic} is used as the default congestion control at the server. 
It is the default congestion control used in the current version of the Linux 
kernel. 

Our measurements are performed on two testbeds: an easily configurable indoor
testbed that emulates different topologies with different characteristics; and
an outdoor testbed using one commercial Internet service provider and a major
cellular carrier in the US. Our outdoor testbed represents real world
scenarios.     

For the scenarios in which we enable MPTCP, we use the stable release (version
v0.86) downloaded from~\cite{MPTCP-UCL}. To provide a fair comparison between
MPTCP and mHTTP, we also use uncoupled congestion control with Cubic for MPTCP.
Uncoupled Cubic represents the case where regular TCP Cubic is used on the
subflows. It increases the size of the congestion window of each subflow irregardless of
the congestion state of the other subflows that are part of the MPTCP session.
We set the maximum receive buffer to $6$MB to avoid potential performance
degradation to MPTCP~\cite{RAICIU12-HHC}. Our testbed configuration is optimized 
for MPTCP. Hence, we observe the best performance we can achieve using MPTCP. 
Our results show that mHTTP performs very close to this baseline.    

We first analyze the overhead of mHTTP and study its effect on the performance
of downloading small objects; we then show the benefits of mHTTP when downloading 
large objects.

\begin{figure}[t]
	\includegraphics[width=\linewidth]{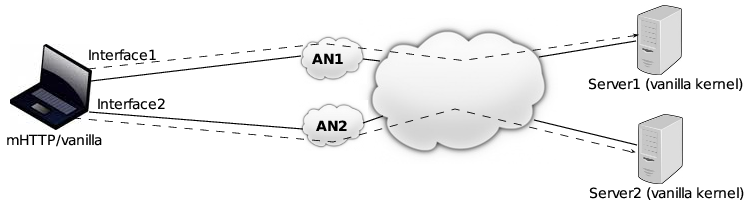}
    \caption{ \label{fig:scenario1} Scenario~1: 2 interfaces at the client; 2 servers; 2 paths (dashed
    lines). In our indoor testbed, AN1 and AN2 are Ethernet routers with a
    nominal rate of $100$Mbps. In our outdoor testbed, the client is a mobile device
    (laptop) with one WiFi and one LTE interface.}
	\vspace*{-0.3cm}
\end{figure}

\subsection{Overhead analysis for small objects}
\label{sec:overhead}
\begin{figure}[t]
\begin{center}
	\includegraphics[width=\linewidth]{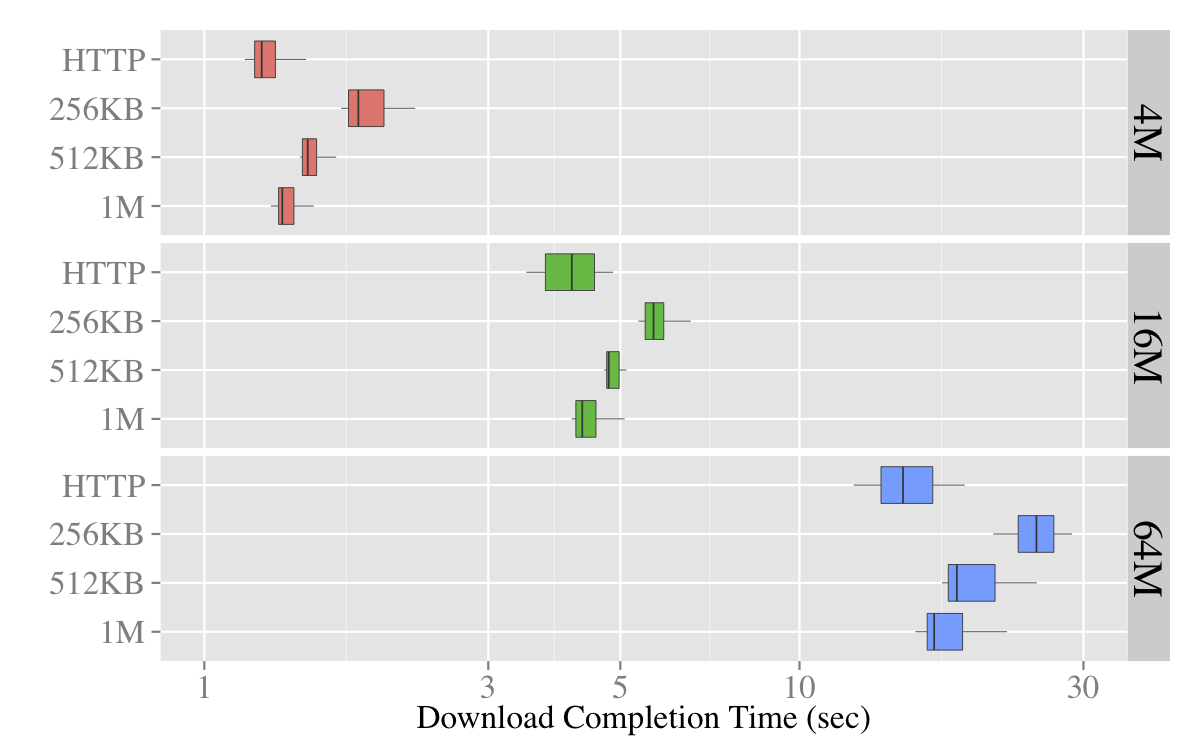}
	\caption{\label{fig:overhead} Overhead analysis: HTTP vs. mHTTP over a single connection.}
\end{center}
	\vspace*{-0.3cm}
\end{figure}

mHTTP suffers a performance degradation each time that a connection performs
a range request. We evaluate this degradation by measuring the download
completion time of a file over a single path connection using HTTP and mHTTP.
The client is connected via an Ethernet interface to an Ethernet router with
a nominal rate of $100$Mbps. The server is also connected to the Ethernet router
via an Ethernet interface. A round trip time of $50$ms of the round-trip time is generated on the
link using a built-in traffic control module of the Linux kernel
(qdisc~\cite{ALMESBERGER99-LNT}). 
We evaluate the overhead of mHTTP with different chunk sizes assuming the
transfer over regular HTTP as the baseline. We show the results in~\fref{fig:overhead}.
We observe that the overhead is more significant with a small chunk
size such as $256$KB than with a large chunk size (e.g. $512$KB, $1024$KB). 
When the chunk size is $1024$KB, we observe that the overhead is around $5-10\%$. 
The poor performance of mHTTP with small chunk sizes is puzzling and a topic for 
future investigation. 

\begin{figure}[t]
\begin{center}
	\includegraphics[width=\linewidth]{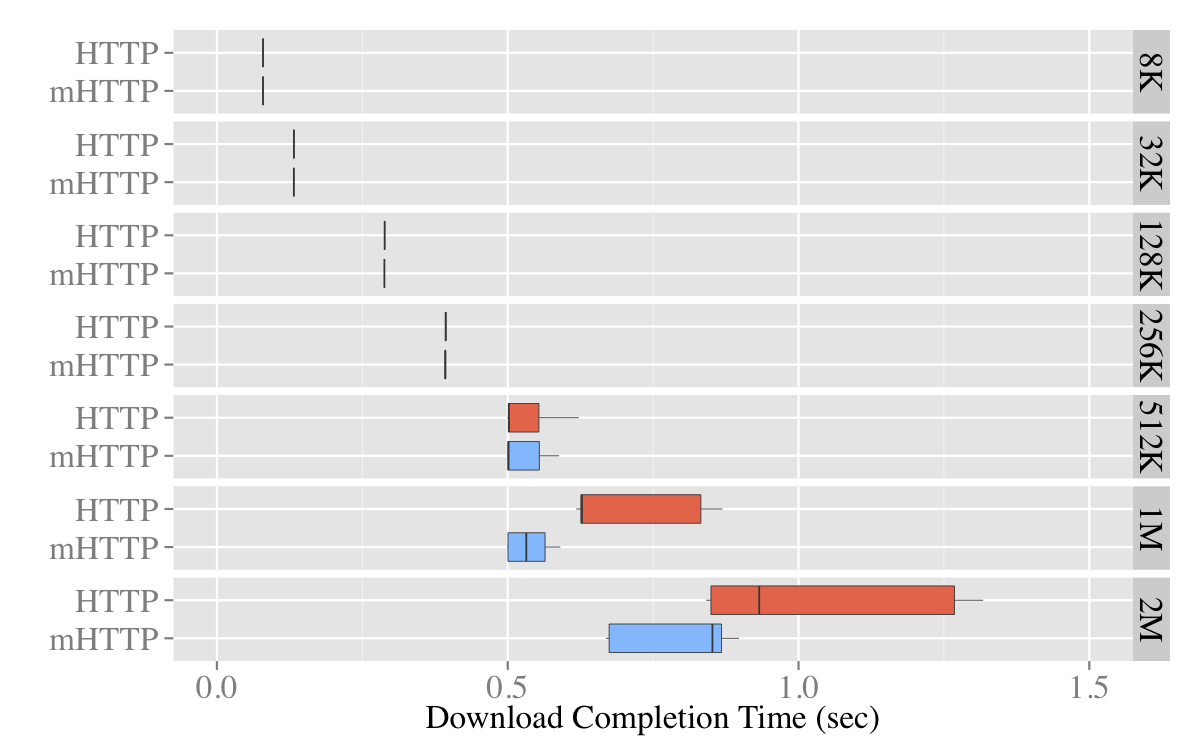}
	\caption{\label{fig:smallflow} The performance of downloading small objects in Scenario~1 (mHTTP chunk size: $512$KB).}
\end{center}
	\vspace*{-0.3cm}
\end{figure}

We now study the performance of mHTTP for downloading small objects. We
evaluate the download completion time of downloading files of various sizes
(from $8$KB to $2$MB) over mHTTP using $512$KB as the chunk size. We consider a
scenario where the client has two interfaces and downloads an object from two
servers as illustrated in~\fref{fig:scenario1}. We emulate AN1 and AN2 with
Ethernet routers with a nominal rate of $100$Mbps. The servers and the client are
connected via Ethernet interfaces to the routers. The round-trip times over the
connections are set to $50$ms. The measurement results are depicted in~\fref{fig:smallflow}.
We observe that mHTTP does not provide any performance gain for small object downloads 
but does no harm either. For object downloads larger than the chunk
size ($512$KB in this measurement), mHTTP provides good performance by utilizing the 
diversity in the network.    

Our results in this section show that mHTTP with large chunk sizes,
such as $512$KB and $1024$KB, provides good performance for
small file downloads and introduces negligible overhead when used
over a single-path connection. In the rest of the paper, we focus our
analysis on the performance of mHTTP for large object downloads.

\begin{figure*}[ht]
\begin{center}
	\includegraphics[width=0.7\linewidth]{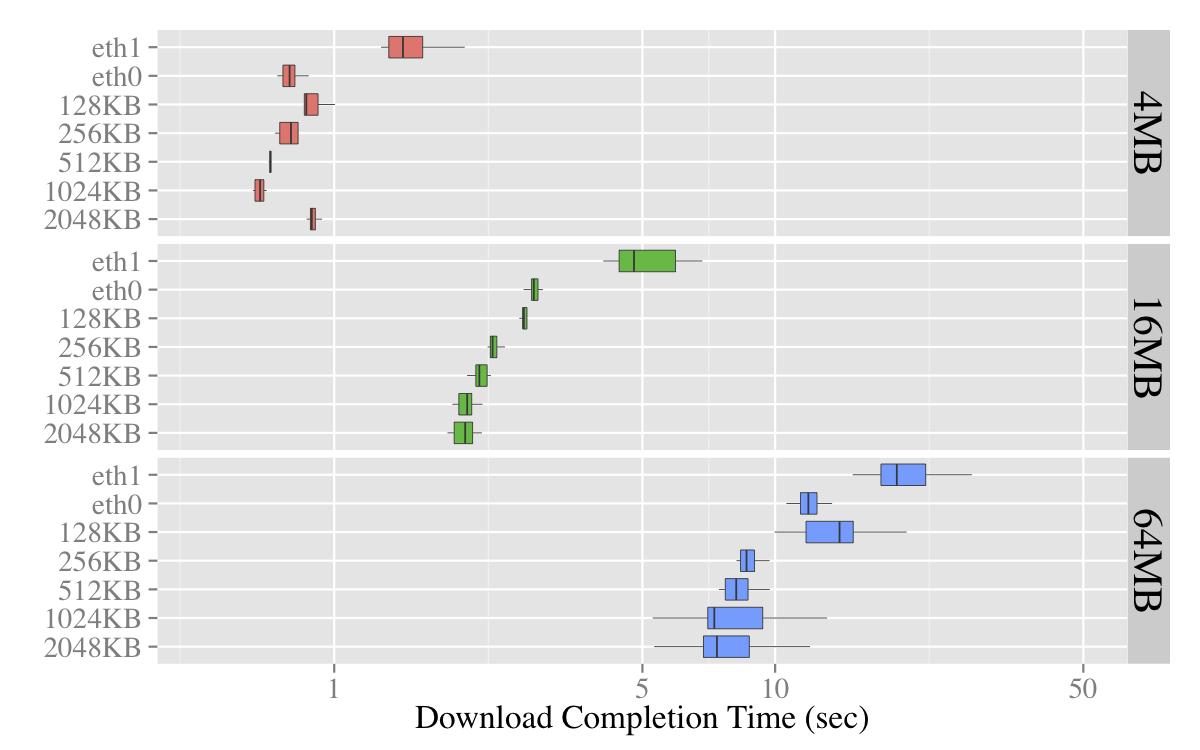}
    \caption{Scenario~1 (indoor testbed): download completion time of regular
    HTTP vs. mHTTP for different file and chunk sizes. mHTTP can efficiently
    use the bandwidth available to the client and outperforms the best
    performing connection among regular HTTP connections. $1024$KB of the chunk
    shows the optimal performance in all file sizes.}
    \label{fig:scenario1-indoor}
\end{center}
\vspace*{-0.3cm}
\end{figure*}

\begin{figure*}[ht]
\begin{center}
	\includegraphics[width=0.7\linewidth]{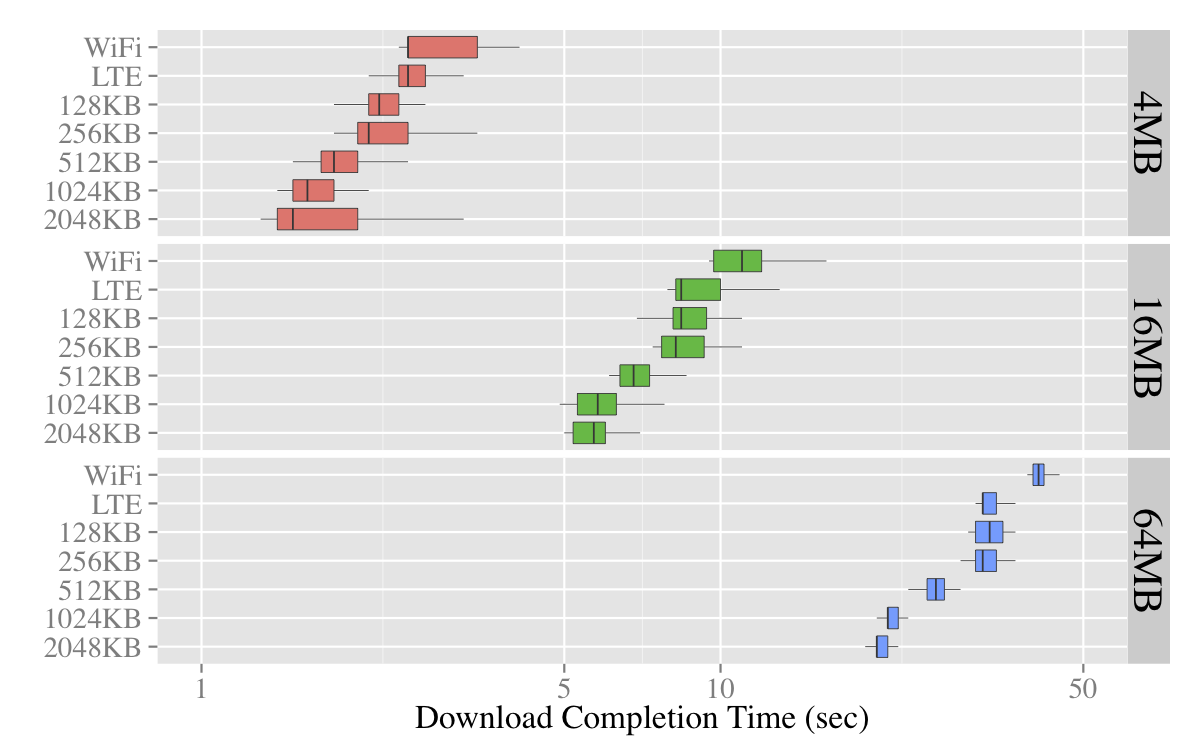}
    \caption{Scenario~1 (outdoor testbed): download completion time of regular
    HTTP vs. mHTTP for different file and chunk sizes. mHTTP can efficiently
    take advantage of the diversity exists in the network. We observe that
    mHTTP shows a relatively low performance when using small chunk sizes
    compared to the measurements with large chunk sizes which is due to the fact that our server
    configuration is not optimized for mHTTP.}
    \label{fig:scenario1-outdoor} 
\end{center}
\vspace*{-0.3cm}	
\end{figure*}

\subsection{mHTTP vs regular HTTP}
\label{sec:mhhtp_vs_sp}
Now, we consider a scenario where the client has multiple interfaces and
downloads a file from multiple servers. We assume a 2-server case in this
scenario. As illustrated in~\fref{fig:scenario1}, the client is equipped with
two interfaces connected to different access networks (ANs). Thus, the client
can establish two different connections to two different servers that contain
identical copies of the same content. Note that MPTCP cannot be used in this
scenario as it is a single-server-oriented protocol.

As the first step, we emulate the above scenario in our indoor testbed where
AN1 and AN2 are Ethernet routers with nominal rates of $100$Mbps each. Each server
is connected via an Ethernet interface to a corresponding router. The client
has two Ethernet interfaces that connect to the routers. In order to emulate
different link latencies in the scenario, we set round-trip times to $10$ms and
$50$ms on the first link and the second link, respectively. The measurement results 
are depicted in~\fref{fig:scenario1-indoor}. We show download completion times 
of file sizes $4$MB, $16$MB, and $64$MB. Each figure compares the performance 
of regular HTTP over a single-path connection with that of mHTTP that uses both 
connections. The results are presented for different mHTTP chunk sizes. 
We observe that (1) the connection over the eth0 interface has a much 
better performance than the one over the eth1 connection; (2) mHTTP greatly benefits
from the existing diversity in the network, the performance gain from using mHTTP 
is larger for larger file sizes; and (3)a chunk size of $1024$KB provides the best 
performance across different file sizes.

\begin{figure}[t]
  \begin{center}
	\begin{minipage}[h]{0.8\linewidth}
		\subfigure[mHTTP buffer size with scheduling.]{\includegraphics[width=\linewidth]{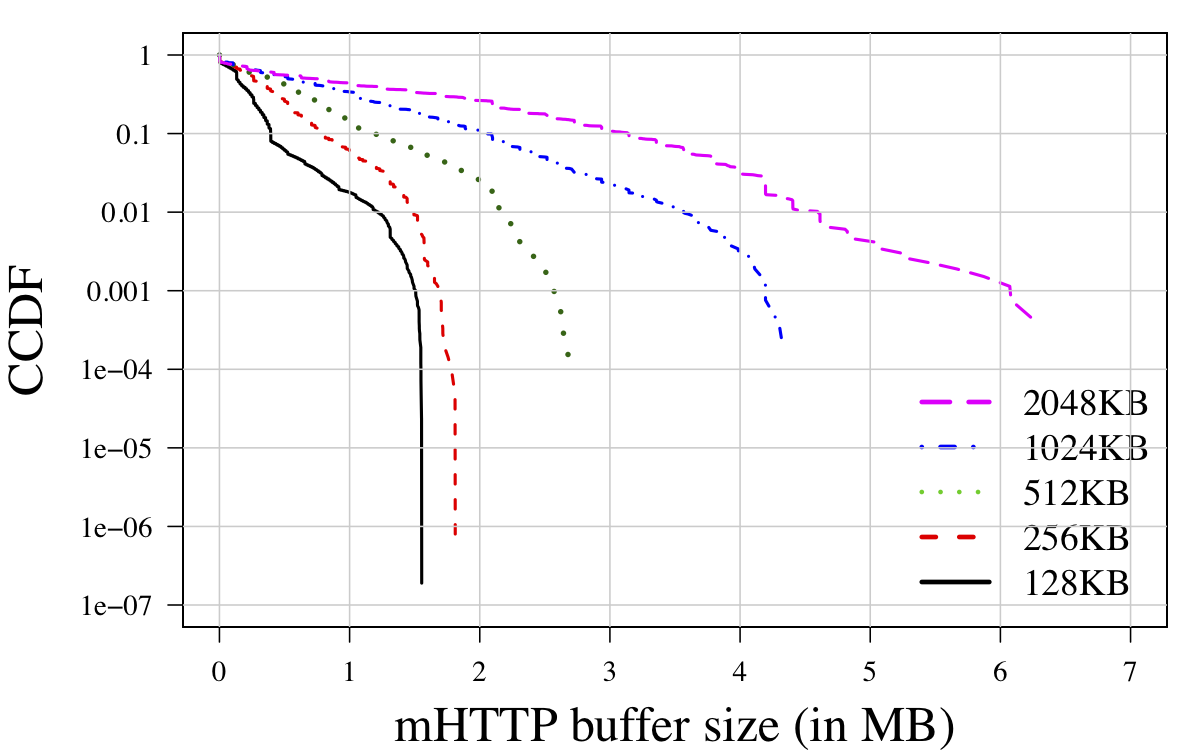}}
	\end{minipage}
	\end{center}
	\hspace{1pt}
	\begin{center}
	\begin{minipage}[h]{0.8\linewidth}
		\subfigure[mHTTP buffer size without scheduling.]{\includegraphics[width=\linewidth]{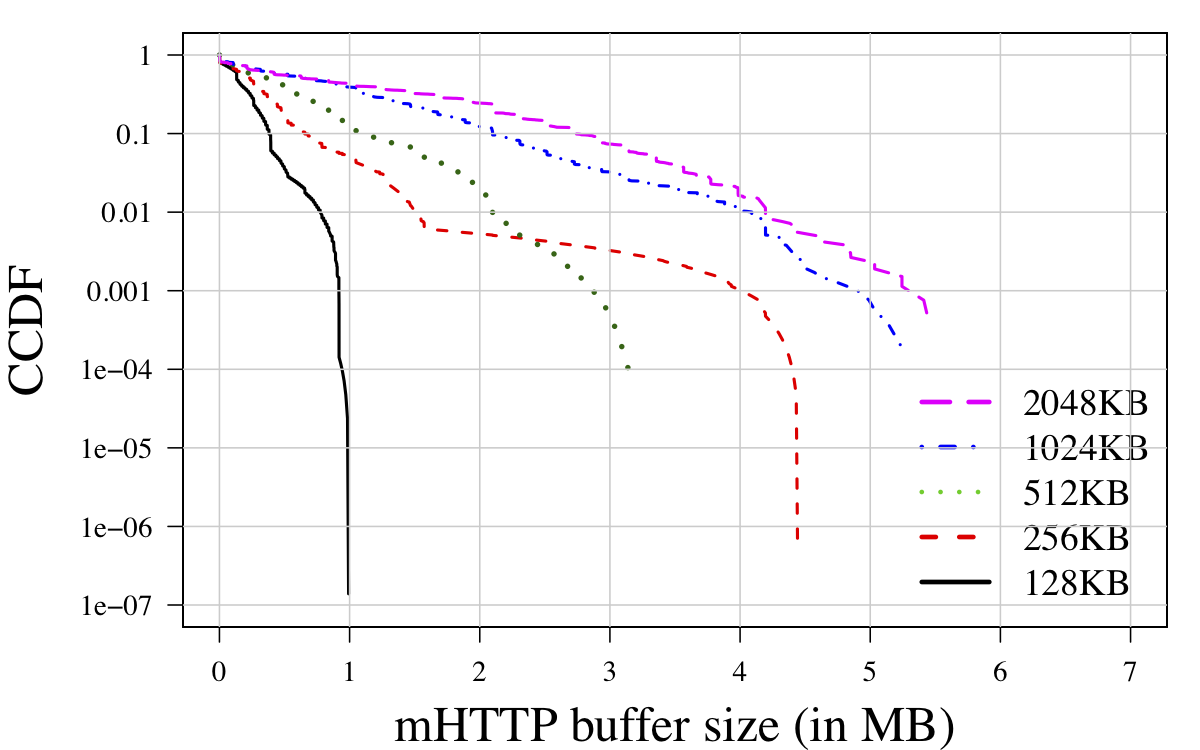}}
	\end{minipage}
	\end{center}
	\caption{\label{fig:scenario1-buffer} Scenario~1: mHTTP buffer size with and without scheduling. Measurements are done in our indoor testbed and for file size of $16$MB.}
  \vspace*{-0.3cm}
\end{figure}

\begin{figure}[t]
\begin{center}
	\includegraphics[width=0.8\linewidth]{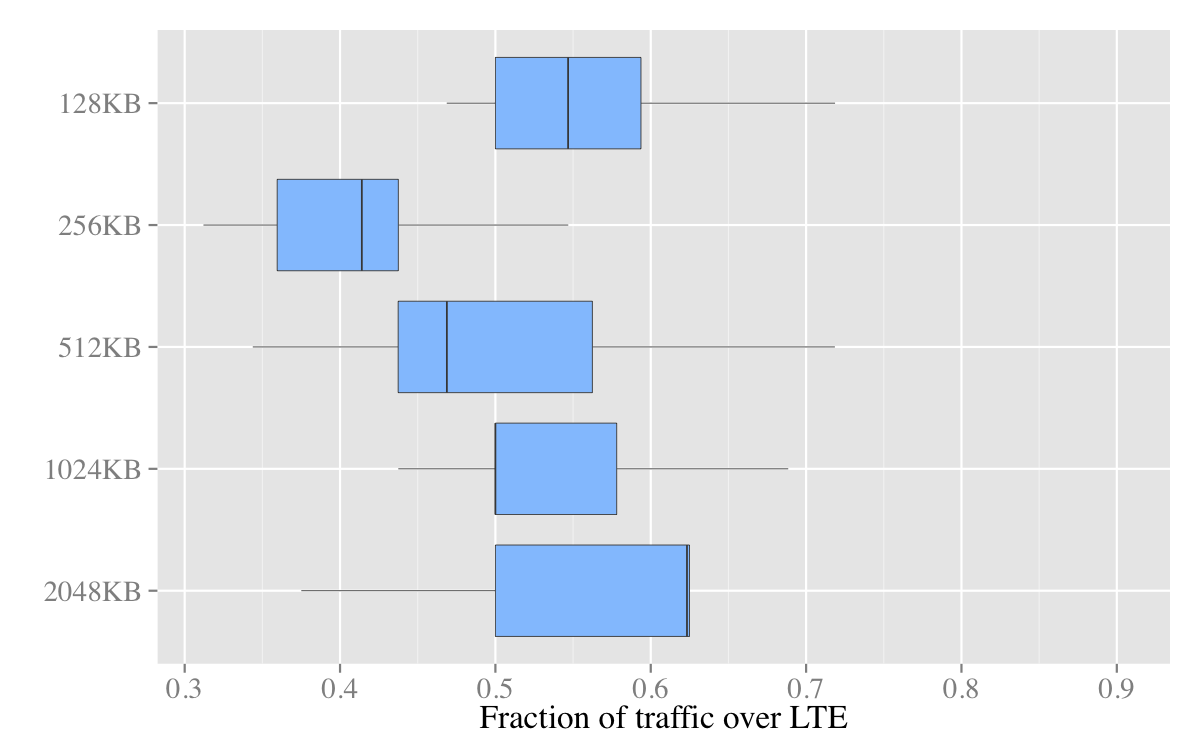}
    \caption{Scenario~1 (outdoor testbed): fraction of traffic carried over a LTE
    connection for file size of $16$MB.}
    \label{fig:scenario1-fraction} 
\end{center}
\vspace*{-0.3cm}	
\end{figure}   

In Section\ref{sec:schedule}, we proposed a scheduler that decides what chunk to request 
over each connection to avoid a bottleneck situation such as presented in~\fref{fig:bottleneck}. 
Here, we compare the performance of this scheduler with a baseline that mHTTP simply requests 
for the next chunk in $D$, whenever it needs to issue a new chunk request over a connection. 
Recall that $D$ is the set of chunks that have not yet been requested for download. 
The experiment is done in our indoor testbed and for file size of $16$MB.

Measurements show that (1) mHTTP with and without scheduler exhibit similar 
performance (in term of download completion time). Hence our scheduler does not affect the 
performance of mHTTP. As the results for mHTTP without scheduling are similar to 
\fref{fig:scenario1-indoor}, we do not show them in this paper. (2) Our scheduler 
efficiently reduces the mHTTP buffer size. \fref{fig:scenario1-buffer} depicts
the CCDF (Complementary Cumulative Distribution Function) of mHTTP buffer sizes for both 
cases. We observe that mHTTP without scheduling requires larger buffer sizes. 
The results for $2048$ chunk size are identical. As in this case we have $8$ chunks to be 
requested over the connections, the scheduler would not have any impact. (3) Furthermore, 
we observe that the mHTTP buffer size is smaller than $1$MB in more than $50\%$ of the cases. 
The maximum buffer occupancy is $7$ MB. Note that mHTTP buffer uses user level memory and not 
the the kernel space memory.

\begin{figure}[t]
  \begin{center}
	\includegraphics[width=\linewidth]{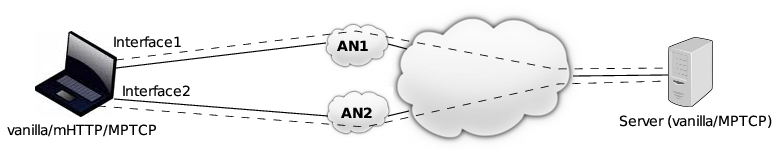}
    \caption{\label{fig:scenario2} Scenario~2: 2 interfaces at the client; 1
    server; 2 paths (dashed lines). As in the previous section, we emulate
    this testbed both indoor and outdoor.}
	\end{center}
  \vspace*{-0.3cm}
\end{figure} 

\begin{figure*}[ht]
\begin{center}
	\includegraphics[width=0.7\linewidth]{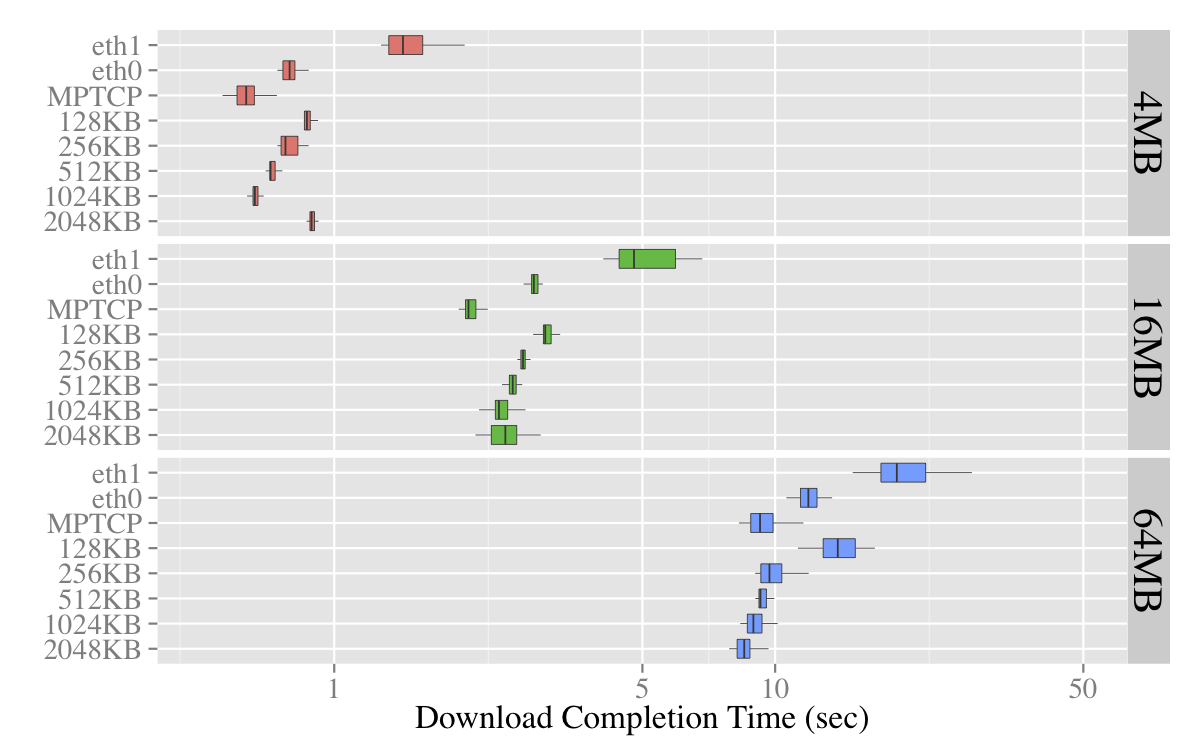}
    \caption{Scenario~2 (indoor testbed):
    download completion time of mHTTP vs. MPTCP and regular HTTP. Our
    testbed configuration is optimized for MPTCP. Hence, MPTCP is able to fully
    utilize the available capacity and provide a good performance. We observe
    that mHTTP perform very close to MPTCP and always outperforms regular HTTP
    over the best path.}
    \label{fig:scenario2-indoor} 
\end{center}
\vspace*{-0.3cm}
\end{figure*}

\begin{figure*}[ht]
\begin{center}
	\includegraphics[width=0.7\linewidth]{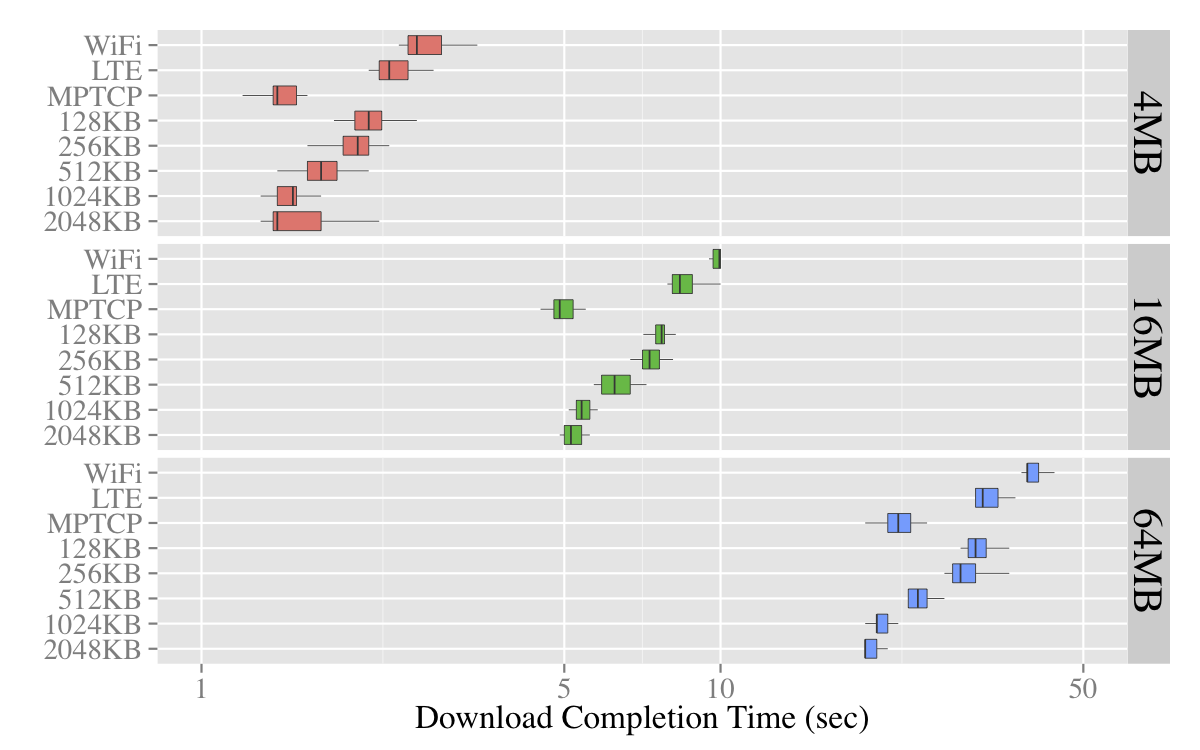}
    \caption{Scenario~2 (outdoor testbed): download completion time of mHTTP
    vs. MPTCP and regular HTTP. We observe that MPTCP is able to fully use
    the available bandwidth and mHTTP performs close to MPTCP.}
    \label{fig:scenario2-outdoor}
\end{center}
\vspace*{-0.3cm}
\end{figure*}

Now, we move our measurements to a more realistic network using our outdoor
testbed. We configure two servers in a university campus and a mobile device
(laptop) as the client. The servers are connected to the Internet via $1$Gbps Ethernet cables 
(i.e., the bottleneck is not at the server). The client device is equipped with two wireless
interfaces (WiFi and LTE) that respectively connect to a WiFi network and a
cellular network. We show the results of our measurements
in~\fref{fig:scenario1-outdoor} for file sizes of $4$MB, $16$MB, and $64$MB and
for different chunk sizes. We observe from the results that (1) LTE and WiFi
exhibit very similar performance; and (2) mHTTP can efficiently use the
available bandwidth, especially when the chunk size is $1024$KB. In this case,
we observe that mHTTP's throughput equals the sum of the throughput of LTE
and WiFi. Hence, mHTTP fully utilizes the available capacity and shows a
substantial performance by reducing the completion time by $50$\perc. For
smaller chunk sizes, we observe a lower performance than that for large chunk
sizes. This is mainly due to the overhead of range requests as analyzed in~\xref{sec:overhead}.
Improving the performance of mHTTP for small chunk sizes is a future research topic.

\fref{fig:scenario1-fraction} depicts the fraction of traffic carried over the
LTE interface using mHTTP. We show the results for different chunk sizes and
for a file size of $16$MB. We observe from~\fref{fig:scenario1-outdoor} that
LTE exhibits a slightly higher throughput than WiFi. Hence, we expect mHTTP to
send more or less the same amount of traffic over LTE and WiFi. Our results
in~\fref{fig:scenario1-fraction} confirm our expectation specifically for
large chunk sizes. 

\subsection{mHTTP vs MPTCP for a single server case}
\label{sec:mhhtp_vs_mptcp}
Our second scenario focuses on comparing the performance of mHTTP and MPTCP in
the multi-homed and single data source environment as illustrated
in~\fref{fig:scenario2}. 

As in the previous section, we first report measurements on our indoor testbed.
The topology of the testbed is slightly changed in this scenario: there is only
one server, and thus no data source diversity. The server and the client are
booted with the MPTCP-enabled kernel when we measure the performance of MPTCP.
The results are shown in~\fref{fig:scenario2-indoor}. As stated before, we
configure our testbed in such a way that it is optimal for MPTCP. Hence, we
expect MPTCP with independent cubic be able to fully utilize the available capacity and provide good
performance. The results confirm this: the MPTCP throughput equals to the sum
of the throughput of two connections. Moreover, we observe that mHTTP performs closely
to MPTCP when the chunk size is $1024$KB.

Furthermore, we observe for $64$MB file size, and for large chunk
sizes, that mHTTP outperforms MPTCP. This is due to the fact that MPTCP uses
a shared TCP receive buffer which can limit its performance when paths have
different characteristics~\cite{RAICIU12-HHC}. However, mHTTP uses a separated
TCP receive buffer for different established connections and hence can perform
well in such a situation.   

Now, we show measurement results from our outdoor testbed: a server
residing  at a university campus and a client equipped with LTE and WiFi
network interfaces. The results are depicted in~\fref{fig:scenario2-indoor}.
Again, we observe that MPTCP fully uses available capacity and
mHTTP performs close to MPTCP, especially for large file sizes and $1024$KB as
the chunk size. 

\begin{figure}[t]
\begin{center}
	\includegraphics[width=0.8\linewidth]{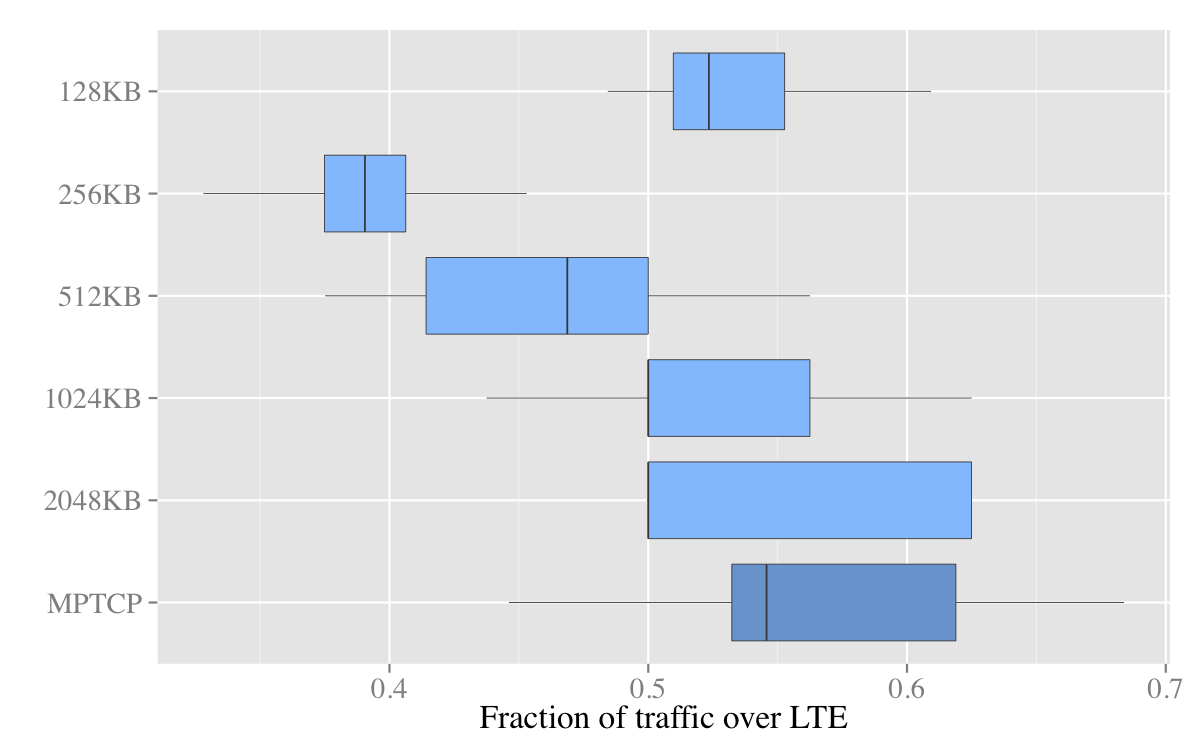}
    \caption{Scenario~2 (outdoor testbed): fraction of traffic carried over a
    LTE connection for $16$MB file using mHTTP as well as MPTCP. We show the results
    for $16$MB file.}
    \label{fig:scenario2-fraction} 
\end{center}
\vspace*{-0.3cm}
\end{figure}

\fref{fig:scenario2-fraction} depicts the fraction of traffic transmitted
over the LTE connection for both mHTTP and MPTCP. We show the results for $16$MB 
file size. As WiFi and LTE connections exhibit similar performance,
we expect that MPTCP and mHTTP transmit more and less the same amount of
traffic over each of these connections as observed in the results. Moreover,
we observe some differences between using different chunk sizes for mHTTP.

\subsection{mHTTP in a multi-source CDN}
\label{sec:real-world}
Finally, we conduct performance measurements on an existing CDN infrastructure. We 
choose a $16$MB file from a well known site on \textit{Alexa.com}'s top-50 list,
where the content is hosted in a CDN. We evaluate the performance of mHTTP when multiDNS 
uses the following two approaches: to simply use Google's public DNS or to leverage separate 
local DNS resolvers.
For the first approach, multiDNS queries Google's DNS over each interface separately, 
and uses the set of IP addresses returned for each interface. For the second approach, 
multiDNS sends a DNS query over each interface to the local DNS of that access network 
to obtain IP addresses. 

\begin{figure}[t]
\begin{center}
	\includegraphics[width=\linewidth]{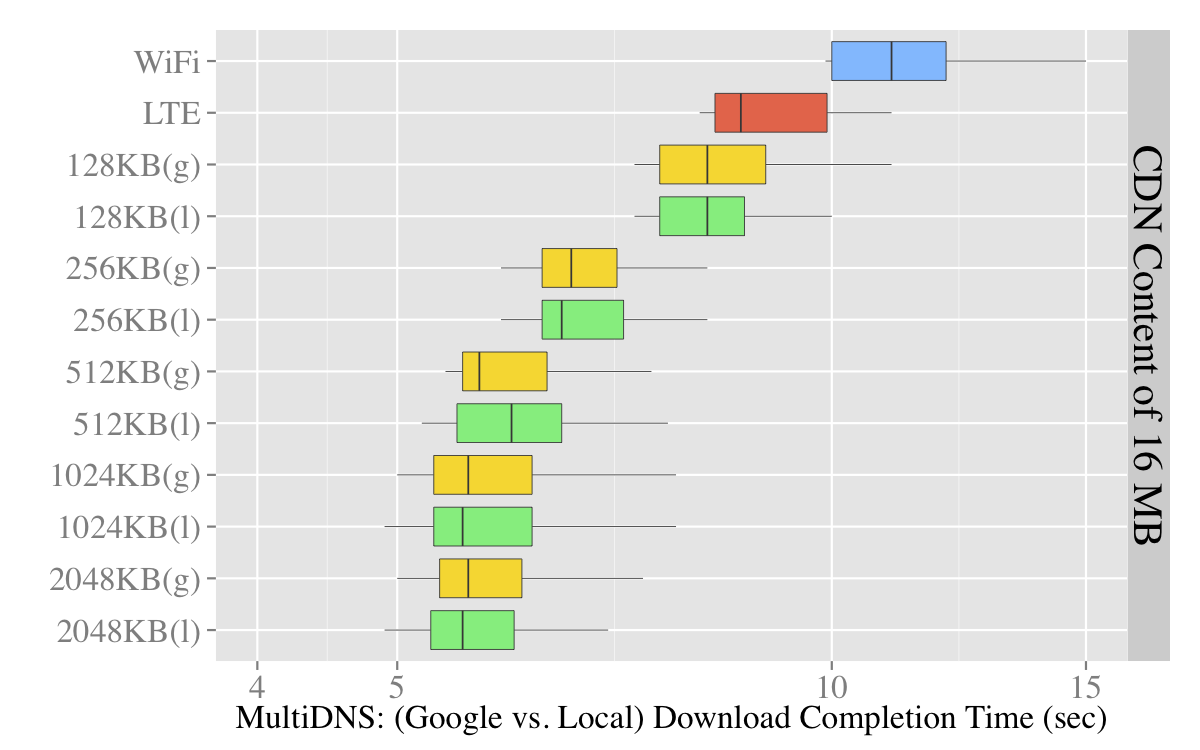}
    \caption{Measurement on a CDN infrastructure. The results are shown for two
    cases: the first case uses IP addresses obtained from DNS 
    queries sent to Google DNS (shown with (g) in the
    y-axis); and the second case that uses IP addresses obtained from the local
    DNS of each of the interfaces networks (shown with (l) in the y-axis).}
    \label{fig:realworld}
\end{center}
\vspace*{-0.3cm}	
\end{figure}

We depict the download times of the file using single-path or mHTTP with different chunk 
sizes in~\fref{fig:realworld}. Note that for single-path TCP, each interface by default 
queries its local DNS resolver. We observe that mHTTP reduces download times by up to 
$50\%$ when compared to the single-path case and performs very well across a wide range 
of chunk sizes. Moreover, no significant differences are observed for both approaches that 
multiDNS uses. Our results in~\fref{fig:realworld} confirms that mHTTP can benefit from 
the path diversity in the Internet and can fully utilize the available bandwidth. 

\subsection{mHTTP is robust to the changes}
Finally, we show through an example how mHTTP performs when one of its 
connections experiences performance drops (due to the congestion either on 
the path or at the server). We use a scenario similar to what is depicted 
in~\fref{fig:scenario1}. We emulate this scenario in our indoor testbed. 
AN1 and AN2 are Ethernet routers with nominal rates of $100$Mbps. Each server is
connected via an Ethernet interface to a corresponding router.
The RTTs of both connections are initially set to 50 ms. The RTT of the second 
connection is configured to be changed to 100ms $3$ seconds after the transfer begins. 
We investigate how mHTTP reacts to this change. We show the results for $64$MB 
file downloads when $1024$KB chunk size is used.

\begin{figure}[t] 
  \begin{center}
	\begin{minipage}[h]{0.9\linewidth}
		\subfigure[\label{fig:timechange}Throughput over each of the connection]{\includegraphics[width=\linewidth]{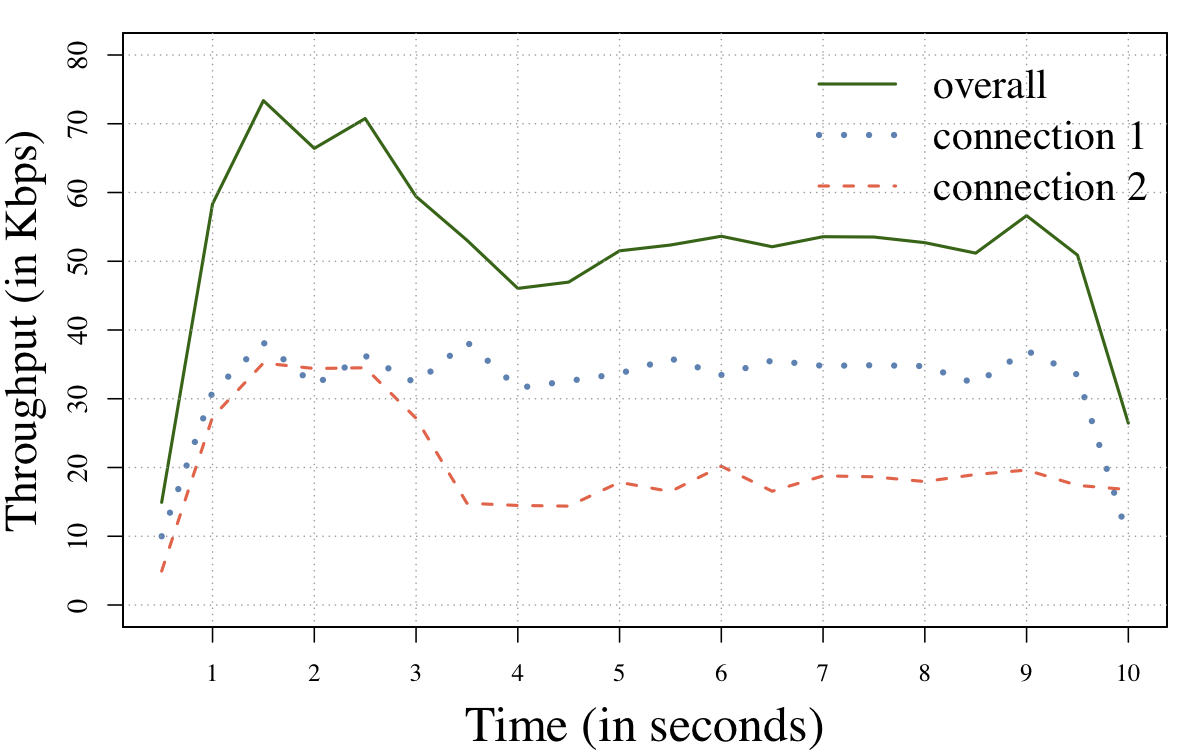}}
	\end{minipage}
	\end{center}
	\hspace{1pt}
	\begin{center}
	\begin{minipage}[h]{0.9\linewidth}
		\subfigure[\label{fig:latencychange}Download completion time]{\includegraphics[width=\linewidth]{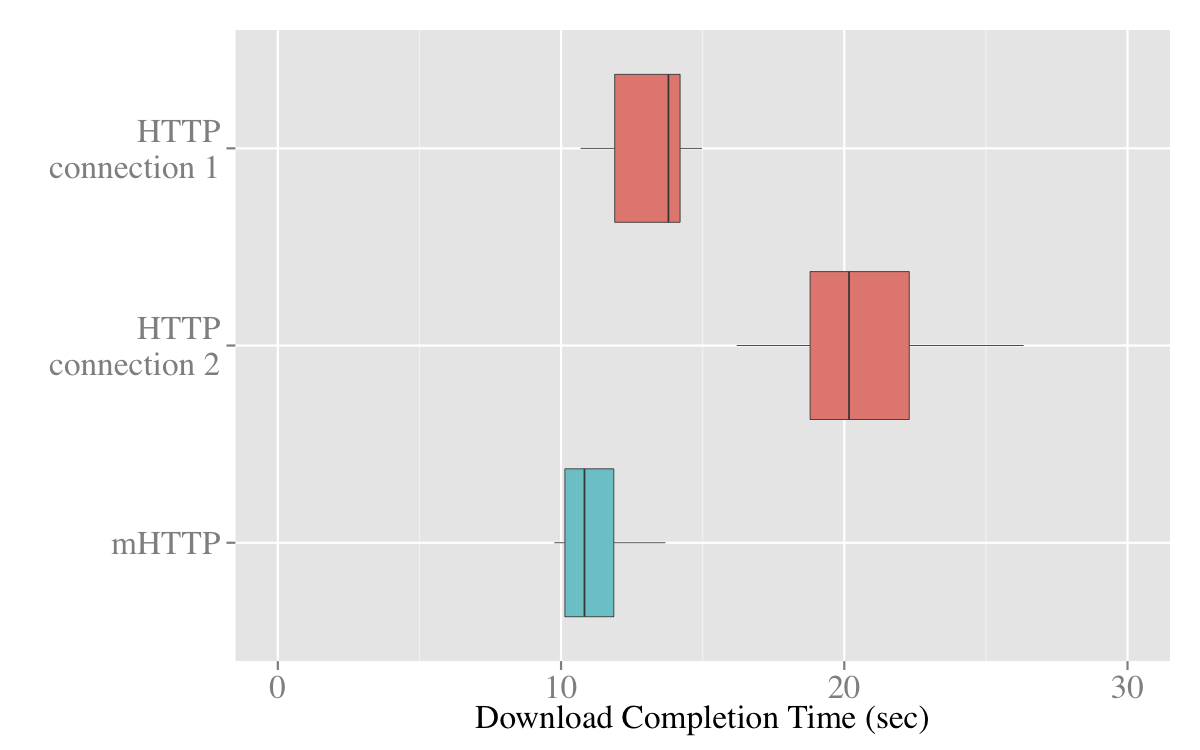}}
	\end{minipage}
	\end{center}
	\caption{The RTT of the second connection is changed to $100$ms after $3$ second. mHTTP is robust to such a latency change and leverages his possible resources over each of these connections.}
  \vspace*{-0.3cm}
\end{figure}

\fref{fig:timechange} depicts the throughput on each of the connection of 
mHTTP and the overall throughput of mHTTP. We show the results for one 
experiment run. We observe that upon the RTT change of the second connection, 
the throughput over this connection decreases. However, mHTTP is robust to such 
a change and takes advantage of the diversity in the network.

\fref{fig:latencychange} depicts the download completion time of mHTTP and compares 
its performance with when we use single-path HTTP over each of these connections 
(recall that the performance of the second connection drop after $3$ second). 
We show the results from $30$ rounds of measurement. We observe that mHTTP provides 
a significant performance gain, especially when we compare it with single-path 
HTTP over the second connection. 
 
\section{Related Work}
\label{sec:related}
The goal of our study is to boost the speed of the HTTP-based content
delivery. Indeed, the need for such a latency reduction in the Internet has already
been acknowledged by network communities.

\textbf{Multipath Approaches}
One of the closest siblings of mHTTP is MPTCP~\cite{RFC-MPTCP,RAICIU12-HHC} which 
is an extension of the regular TCP that enables a user to spread its traffic across 
disjoint paths. Although MPTCP
focuses on the path diversity between a single server and a single receiver and
requires the modification at both end-hosts, the fundamental idea behind these two
protocols is the same. Furthermore, mHTTP sheds light on solving a middlebox
conundrum~\cite{OLIVIER-MPTCP} that MPTCP currently struggles with.
Kaspar~\cite{KASPA12-PHD} thoroughly studies the path diversity in the Internet and discusses
use cases on the transport layer as well as on the application layer. His work
and mHTTP have many features in common except that his work is limited on a 
single client/server scenario and it does not take scheduling into the design consideration.

\textbf{Multi-source Approaches}
Content Distribution Network (CDN) is a key technology for reducing the
delivery latency in today's Internet and the performance of CDN has been
evaluated by many 
studies~\cite{HUANG08-MEL,KARAGIANNIS05-SIS,KRISHNAMURTHY01-UPC,RATNASAMY02-TOC}.
CDNs provide widely distributed servers with multiple copies of the content
available at different locations. CDN typically selects the content server
based on the IP address of client's DNS server and it often makes an incorrect
suggestion due to the use of a public DNS~\cite{AGER10-CDR} or the
malfunction of the IP geolocation database~\cite{POESE11-IGD}. To improve the
performance of the server selection, mechanisms such as
PaDIS~\cite{POESE10-ICD} or ALTO~\cite{RFC-ALTO} have been discussed in the community.
Our goal is to leverage such content distribution infrastructures. 
However, mHTTP does not limit the communication to a single server. Instead, mHTTP
connects to multiple content servers for utilizing path and server diversity of
the Internet. Tian~\etal~\cite{TIAN12-TAA} has proposed a mechanism that is an 
extension of DASH~\cite{ADHIKARI-UNU} video streaming protocol, but heavily 
relies on specific features of DASH. mHTTP, on the other hand, can be used for any
HTTP-type traffic, including streaming contents. 

\textbf{Single Path Approaches}
Google has proposed a new protocol, SPDY~\cite{BELSHE12-SPDY,URL-SPDY},
which shares the ultimate goal with mHTTP, i.e., reducing the user latency. 
These two protocols (mHTTP and SPDY) have a similar architecture that
does not need any modifications in existing applications. 
SPDY uses only one server and one interface at a time and utilizes a single TCP 
connections as if there are multiple connections in it.
To achieve this, server-side socket APIs must support SPDY and that clearly
distinguishes SPDY from mHTTP. However, features of mHTTP and SPDY are mutually
exclusive, thus these two protocols may even be merged in the same platform.

\textbf{Application Specific Approaches}
Bittorrent implements a sophisticated mechanism which enables users to download
the same content from multiple sources~\cite{COHEN03-IBR}. However, Bittorrent is an
application specific protocol and it needs modification both at the sender's and at
the receiver's side. Download managers, often run as add-on software in a web
browser or as stand-alone software, can be other examples of application
specific approaches,~\eg~\cite{URL-FLASH,URL-JDOWN}.


\section{Summary}
\label{sec:discussion}
Advantages of simultaneously utilizing multiple paths over a network
communication are widely evaluated and 
understood~\cite{BARRE11-MTF,CHEN13-MSM,NGUYEN11-EMT,RAICIU11-IDP}. Given the
fact that HTTP accounts for more than 60\perc~\cite{MAIER09-ODC} of today's
Internet traffic and that the major fraction of the total web servers are
operated on content distribution infrastructures~\cite{HUANG08-MEL,TRIUKOSE09-CDN}, 
it is meaningful to broaden the benefit via globally replicated content sources. 
However, convincing application developers and content providers to modify/update 
their software is practically infeasible within a reasonable amount of time. 
mHTTP's key contribution is to bring significant benefits to the end-to-end content
delivery by utilizing the path diversity in the Internet without any changes on
existing applications and the server-side network stack. 

Our results show that the performance gain of mHTTP is relatively lower when
small chunk sizes are used. Part of the problem is due to our testbed configuration.
Additionally, our implementation is still in the testing  phase. Optimizing the HTTP 
parser and restructuring mHTTP buffer will provide substantial performance increase. 
This is a topic for future investigation. 

For small object downloads, we observed that mHTTP does not provide a high
performance gain, but does not harm either. Moreover, we can modify mHTTP such
that it does not establish multiple paths if the object size is relatively
small. Hence, for the small flows, mHTTP will fall back to regular HTTP.
Furthermore, we can integrate ideas like socket intense proposed
in~\cite{socket-intense} in our implementation to better deal with small flows.  

In regard to the comparison with MPTCP in single server scenarios, we
observe that mHTTP exhibits similar performance as MPTCP for large chunk sizes,
e.g., $1024$KB, and for downloading large objects. Moreover, previous studies
show that MPTCP, similarly to mHTTP, does not provide a high performance gain
for small object downloads~\cite{CHEN13-MSM}. Hence, we consider mHTTP to be 
a viable alternative when running HTTP. Note that MPTCP requires changes to the 
kernel, both at the sender and receiver. mHTTP, on the other hand, requires only 
receiver-side modifications which are restricted to the socket interface.  

In our current mHTTP implementation, we do not allow a user to establish more
than two connections and, in particular, not more than one connection over an
interface. However, we let the user leverage his possible resources, that is decided
by regular TCP, over each of these connections. Hence, an mHTTP user will not
be more aggressive than a TCP user over each interface and will not use the
network bandwidth more than twice than a regular single-path user. This provides
some level of fairness in the network. 

We can use similar mechanisms as MPTCP, coupled-control~\cite{mptcprfc} or
OLIA~\cite{KHALILI12-MIN}, to provide load balancing across multiple connections of a
content download. As our design goal is not to modify servers, this can be
implemented by modifying the TCP kernel of the receiver. The idea is that we
can limit/regulate the transmission rate over a connection by adjusting the
receive window size advertised by the receiver. This is a topic for future research.

Finally, We plan to extend our study to other use cases such as a streaming
content delivery (e.g., YouTube and/or Netflix) from multiple data sources.
Specifically, we are interested in studying if using mHTTP can reduce the
start-up latency of streaming contents~\cite{RAO11-NCO}. The performance study 
of mHTTP in high Bandwidth-Delay-Product environments is another future research
topic.



\section{Acknowledgements}
This work was supported by the EU project CHANGE (FP7-ICT-257422) and by the EIT KIC
project MONC. This material is also based upon work supported by the US Army Research 
laboratory and the UK Ministry of Defense under AgreementNumber W911NF-06-3-0001 
and by the National Science Foundation under Grants IIS-0916726 and CNS-1040781.

\bibliographystyle{acm}
\bibliography{main}

\end{document}